\documentclass[twoside,twocolumn,9pt]{article}
\usepackage{extsizes}
\usepackage[super,sort&compress,comma]{natbib} 
\usepackage[version=3]{mhchem}
\usepackage[left=1.5cm, right=1.5cm, top=1.785cm, bottom=2.0cm]{geometry}
\usepackage{balance}
\usepackage{mathptmx}
\usepackage{sectsty}
\usepackage{graphicx} 
\usepackage{lastpage}
\usepackage[format=plain,justification=justified,singlelinecheck=false,font={stretch=1.125,small,sf},labelfont=bf,labelsep=space]{caption}
\usepackage{float}
\usepackage{fancyhdr}
\usepackage{fnpos}
\usepackage[english]{babel}
\addto{\captionsenglish}{%
  
}
\usepackage{array}
\usepackage{droidsans}
\usepackage{charter}
\usepackage[T1]{fontenc}
\usepackage[usenames,dvipsnames]{xcolor}
\usepackage{setspace}
\usepackage[compact]{titlesec}
\usepackage{hyperref}

\usepackage{multirow,makecell}
\usepackage{amsmath}
\usepackage{dcolumn}
\usepackage{bm}
\usepackage[normalem]{ulem}  
\usepackage{breqn}

\usepackage{changes}

\newcommand{\di}{D}

\definecolor{cream}{RGB}{222,217,201}

\makeatletter
\@namedef{Changes@AuthorColor}{red}
\colorlet{Changes@Color}{red}
\makeatother

\begin{document}

\pagestyle{fancy}
\thispagestyle{plain}
\fancypagestyle{plain}{
\renewcommand{\headrulewidth}{0pt}
}

\makeFNbottom
\makeatletter
\renewcommand\LARGE{\@setfontsize\LARGE{15pt}{17}}
\renewcommand\Large{\@setfontsize\Large{12pt}{14}}
\renewcommand\large{\@setfontsize\large{10pt}{12}}
\renewcommand\footnotesize{\@setfontsize\footnotesize{7pt}{10}}
\makeatother

\renewcommand{\thefootnote}{\fnsymbol{footnote}}
\renewcommand\footnoterule{\vspace*{1pt}%
\color{cream}\hrule width 3.5in height 0.4pt \color{black}\vspace*{5pt}} 
\setcounter{secnumdepth}{5}

\makeatletter 
\renewcommand\@biblabel[1]{#1}            
\renewcommand\@makefntext[1]%
{\noindent\makebox[0pt][r]{\@thefnmark\,}#1}
\makeatother 
\renewcommand{\figurename}{\small{Fig.}~}
\sectionfont{\sffamily\Large}
\subsectionfont{\normalsize}
\subsubsectionfont{\bf}
\setstretch{1.125} 
\setlength{\skip\footins}{0.8cm}
\setlength{\footnotesep}{0.25cm}
\setlength{\jot}{10pt}
\titlespacing*{\section}{0pt}{4pt}{4pt}
\titlespacing*{\subsection}{0pt}{15pt}{1pt}

\fancyfoot{}
\fancyfoot[LO,RE]{\vspace{-7.1pt}\includegraphics[height=9pt]{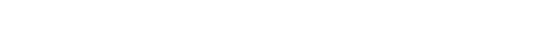}}
\fancyfoot[CO]{\vspace{-7.1pt}\hspace{13.2cm}\includegraphics{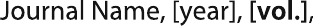}}
\fancyfoot[CE]{\vspace{-7.2pt}\hspace{-14.2cm}\includegraphics{head_foot/RF}}
\fancyfoot[RO]{\footnotesize{\sffamily{1--\pageref{LastPage} ~\textbar  \hspace{2pt}\thepage}}}
\fancyfoot[LE]{\footnotesize{\sffamily{\thepage~\textbar\hspace{3.45cm} 1--\pageref{LastPage}}}}
\fancyhead{}
\renewcommand{\headrulewidth}{0pt} 
\renewcommand{\footrulewidth}{0pt}
\setlength{\arrayrulewidth}{1pt}
\setlength{\columnsep}{6.5mm}
\setlength\bibsep{1pt}

\makeatletter 
\newlength{\figrulesep} 
\setlength{\figrulesep}{0.5\textfloatsep} 

\newcommand{\topfigrule}{\vspace*{-1pt}%
\noindent{\color{cream}\rule[-\figrulesep]{\columnwidth}{1.5pt}} }

\newcommand{\botfigrule}{\vspace*{-2pt}%
\noindent{\color{cream}\rule[\figrulesep]{\columnwidth}{1.5pt}} }

\newcommand{\dblfigrule}{\vspace*{-1pt}%
\noindent{\color{cream}\rule[-\figrulesep]{\textwidth}{1.5pt}} }

\makeatother

\twocolumn[
  \begin{@twocolumnfalse}
{\includegraphics[height=30pt]{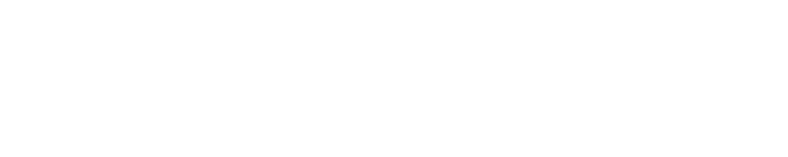}\hfill\raisebox{0pt}[0pt][0pt]{\includegraphics[height=55pt]{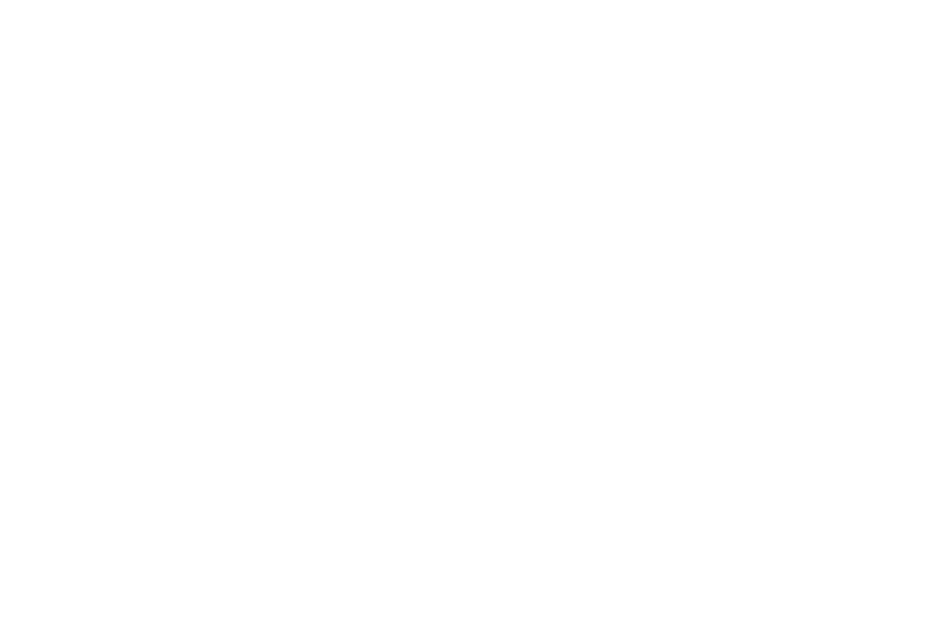}}\\[1ex]
\includegraphics[width=18.5cm]{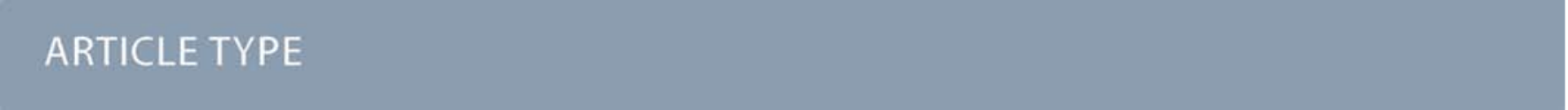}}\par
\vspace{1em}
\sffamily
\begin{tabular}{m{4.5cm} p{13.5cm} }

\includegraphics{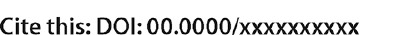} & \noindent\LARGE{\textbf{Stress propagation in locally loaded packings of disks and pentagons$^\dag$}} \\
\vspace{0.3cm} & \vspace{0.3cm} \\

 & \noindent\large{Ryan Kozlowski,$^{\ast}$\textit{$^{a}$} Hu Zheng,\textit{$^{b}$} Karen E. Daniels,\textit{$^{c}$} and Joshua E. S. Socolar\textit{$^{a}$}} \\

\includegraphics{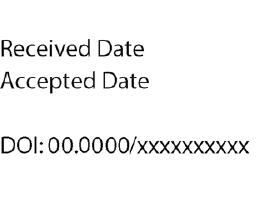} & \noindent\normalsize{The mechanical strength and flow of granular materials can depend strongly on the shapes of individual grains.  We report quantitative results obtained from photoelasticimetry experiments on locally loaded, quasi-two-dimensional granular packings of either disks or pentagons exhibiting stick-slip dynamics. Packings of pentagons resist the intruder at significantly lower packing fractions than packings of disks, transmitting stresses from the intruder to the boundaries over a smaller spatial extent. Moreover,  packings of pentagons feature significantly fewer back-bending force chains than packings of disks. Data obtained  on the forward spatial extent of stresses and back-bending force chains collapse when the packing fraction is rescaled according to the packing fraction of steady state open channel formation, though data on intruder forces and dynamics do not collapse. We  comment on the influence of system size on these findings and highlight connections with the dynamics of the disks and pentagons during slip events. 
} 

\end{tabular}

 \end{@twocolumnfalse} \vspace{0.6cm}

  ]

\renewcommand*\rmdefault{bch}\normalfont\upshape
\rmfamily
\section*{}
\vspace{-1cm}


\footnotetext{\textit{$^{a}$~Department of Physics, Duke University, Durham, North Carolina 27708, USA. E-mail: rhk11@phy.duke.edu}}
\footnotetext{\textit{$^{b}$~Department of Geotechnical Engineering, College of Civil Engineering, Tongji University, Shanghai 200092, China. }}
\footnotetext{\textit{$^{c}$~Department of Physics, North Carolina State University, Raleigh, North Carolina 27695, USA. }}

\footnotetext{\dag~Electronic Supplementary Information (ESI) available:  }


\date{\today}

\section{Introduction}
\label{sec:intro}

Observations of the mechanical response of a granular medium to a local point load highlight connections among grain-scale mechanics and dynamics, mesoscopic force propagation, and bulk mechanics \cite{sticksliptordesillas,probeintruderreichhardt,slowdraggeng,pointloadexperimentsatmangeng, Geng2003greensfunction, pointloadsimulationgoldenberg}. Many simulations and experiments have probed the dynamics of local intruding rods and grains and the corresponding flow of grains \cite{dragforcecavityformationintruderkolb,rheologyintruderexperimentseguin,intrudervibrationgdauchot,probeintruderreichhardt,sticksliptordesillas}. These efforts have revealed how driving parameters such as loading rate \cite{dragforcecavityformationintruderkolb,rheologyintruderexperimentseguin} or driving spring stiffness \cite{stickslipalbert,sticksliptordesillas} as well as system characteristics including system size \cite{dragforcecavityformationintruderkolb, stickslipalbert}, friction \cite{KozlowskiBasalFrictionPRE,KozlowskiCarlevaroSimulationBasalFrictionPoint}, and packing fraction \cite{dragforcecavityformationintruderkolb,intrudervibrationgdauchot,slowdraggeng,KozlowskiBasalFrictionPRE} influence the mechanical response of a packing. 
In the present work, we explore the role that \textit{grain shape} has on the mechanical response and dynamics of a packing to an intruder. To do so, we drive a grain-sized intruder through packings composed of either pentagons or disks to compare the features of stress transmission through granular materials with and without angularity.

Spherical (circular in 2D) grains are the simplest to model computationally and analyze experimentally, and the parameter space available to describe non-spherical shapes is infinite. Yet isolating crucial shape differences between spherical grains and other shapes is necessary for characterizing the mechanics and dynamics of real granular materials found in geophysical systems \cite{Marone2002ParticleShapeInGranularFault,Anthony2005ShapeInReal,ChoSantamarina2006SandStrength}, processed in pharmaceutical and agricultural industries, and utilized in civil engineering \cite{Dierichs2015ArchitectureConvexity}. In the past two decades, advances in technologies that can mass-produce grains of well defined shapes (e.g., 3D printing \cite{JaegarKieranShape,Athanassiadis20143DprintGrainShape} and laser cutting \cite{ZhuTang2018dissertation,fazelpour2021effect}) and computational tools that can efficiently model aspherical grains \cite{Azema2012Angularity,pytlos2015modelling,Kawamoto2018ShapeLSDEM,Zhao2019TetrahedralGMDEM} have enabled numerous studies of granular materials with nonspherical grains. Many studies have characterized the mechanics and dynamics of elongated grains, such as plates or rods with varying aspect ratio \cite{Borzsonyi2013AnisotropicGrainReview,experimentswithellipsoids,ellipsoidaljamnagelliu,Tang2016ellipsesHopper,Donevellipsoidsm&M,AzemaRadjaiElongated,Zuriguel2007stressdipsellipses,Farhadi2014ellipses,Teitel2018Spherocylinders,Wegner2014ShapeDilatancy,Ashour2017AnisotropicGrainHopper,Chen2021SoftEllipsoidalPacking,Schreck2010DimerEllipses,GonzalezPinto2017VibratedMonolayersRods}; grains with concavities, such as staple-like grains or hexapods \cite{murphystaplesZcolumns,yuchenhexapods,Gravish2012StapleEntangle,Szarf2011NonconvexDiscClumps,Athanassiadis20143DprintGrainShape,JaegarKieranShape}; and angular grains, such as tetrahedra or pentagons \cite{HyperstaticTetrahedra,JaegarKieranShape,Athanassiadis20143DprintGrainShape,ShapeFrictionAzema,yiqiushape,Jaoshvili2010RandomPackingTetrahedra,IrregularPolyhedralPacking,Wegner2014ShapeDilatancy,Zhao2019TetrahedralGMDEM}. 

Grain-scale interactions between nonspherical grains can differ significantly from the interactions that can arise between spherical grains. For instance, the presence of flat edges or significant elongation of grains introduces rotational resistance for individual grains experiencing force-bearing edge-edge (or face-face) contacts \cite{Estrada2011rollingresistanceAngularity,HyperstaticTetrahedra,IrregularPolyhedralPacking,ShapeFrictionAzema,yiqiushape,Zhao2019TetrahedralGMDEM}, and concave particles can interlock and support tensile stresses \cite{Gravish2012StapleEntangle,yuchenhexapods}. 
Microscopic interactions associated with various grain shapes have been shown to influence macroscopic properties of granular materials in many contexts, including the shear strength  \cite{ShapeFrictionAzema,JaegarKieranShape,Athanassiadis20143DprintGrainShape,Marone2002ParticleShapeInGranularFault,yuchenhexapods} and  structural characteristics of jammed packings \cite{yiqiushape,Donevellipsoidsm&M,Zhao2019TetrahedralGMDEM,ellipsoidaljamnagelliu,Zuriguel2007stressdipsellipses,pentagonalforcetransmissionazema,KondicPersistenceDisksPents1,Zuriguel2007stressdipsellipses,Wang2015StructuralCharacteristicsRegPolygons,Santos2020TwistingFriction,Schreck2010DimerEllipses,Harrington2019StructuralDefectsDimersEllipses,XuYiqiu2017HeptPentDiskJam,HuZheng2017CrossJam,Teich2016ClustersPolyhedra,Brito11736,Wegner2014ShapeDilatancy,Chen2021SoftEllipsoidalPacking}, the dynamical organization of anisotropic grains in driven granular media \cite{KaiHuang2017HexagonRotatorGM,Farhadi2014ellipses,Tang2016ellipsesHopper,AzemaRadjaiElongated,GonzalezPinto2017VibratedMonolayersRods}, the rheology of dense granular flows \cite{Salerno2018ShapeFrictionPackingFlow,ZhuTang2018dissertation,fazelpour2021effect}, the flow and clogging dynamics of granular materials driven through restricted apertures \cite{Tang2016ellipsesHopper,Goldberg_Carlevaro_Pugnaloni_shapeclogging,Ashour2017AnisotropicGrainHopper,CloggingWithPentsDiskSims,hafezcloggingshape}, 
and the shear thickening of sheared suspensions \cite{Singh2020RollOfFriction}. 

In recent work (Ref.~\cite{PGKoz2021}) we experimentally investigated the influence of grain angularity on the irregular stick-slip dynamics \cite{baumbergerstickslipcreep,stickslipnasuno,stickslipalbert} of a spring-driven, single-grain intruder pushed through quasi-2D monolayers of either cylinders (disk-packings) or regular pentagonal prisms (pentagon-packings). We showed that the average force exerted on the intruder by a pentagon-packing is comparable with the force exerted by a substantially more dense disk-packing. We also observed that pentagonal grains rotate and circulate around the intruder far less during slip events than do disks. In the present paper, we further explore differences in the disk- and pentagon-packings by using photoelasticimetry to quantify differences in the transmission of forces within the granular material during stable sticking periods. 

We find that (1) stresses propagate significantly further in front of the intruder in disk-packings as compared with pentagon-packings given comparable loading forces, and the spatial extent of stress networks is comparable to the size of the dense cluster of grains in front of the intruder, and (2) force chains bend behind the intruder more frequently, and with greater spatial extent, for disks than for pentagons. Both of these measures for pentagons and disks collapse onto a universal trend when the packing fraction $\phi$ is rescaled by $\phi_0$, the packing fraction at which the material forms an open channel through which the intruder flows unimpeded in the steady state. 

\section{Experimental methods}
\label{sec:expmethods}

\subsection{2D granular system with spring driven intruder}
\label{ssec:system}
\begin{figure}
    \centering
    \includegraphics[width=1\linewidth]{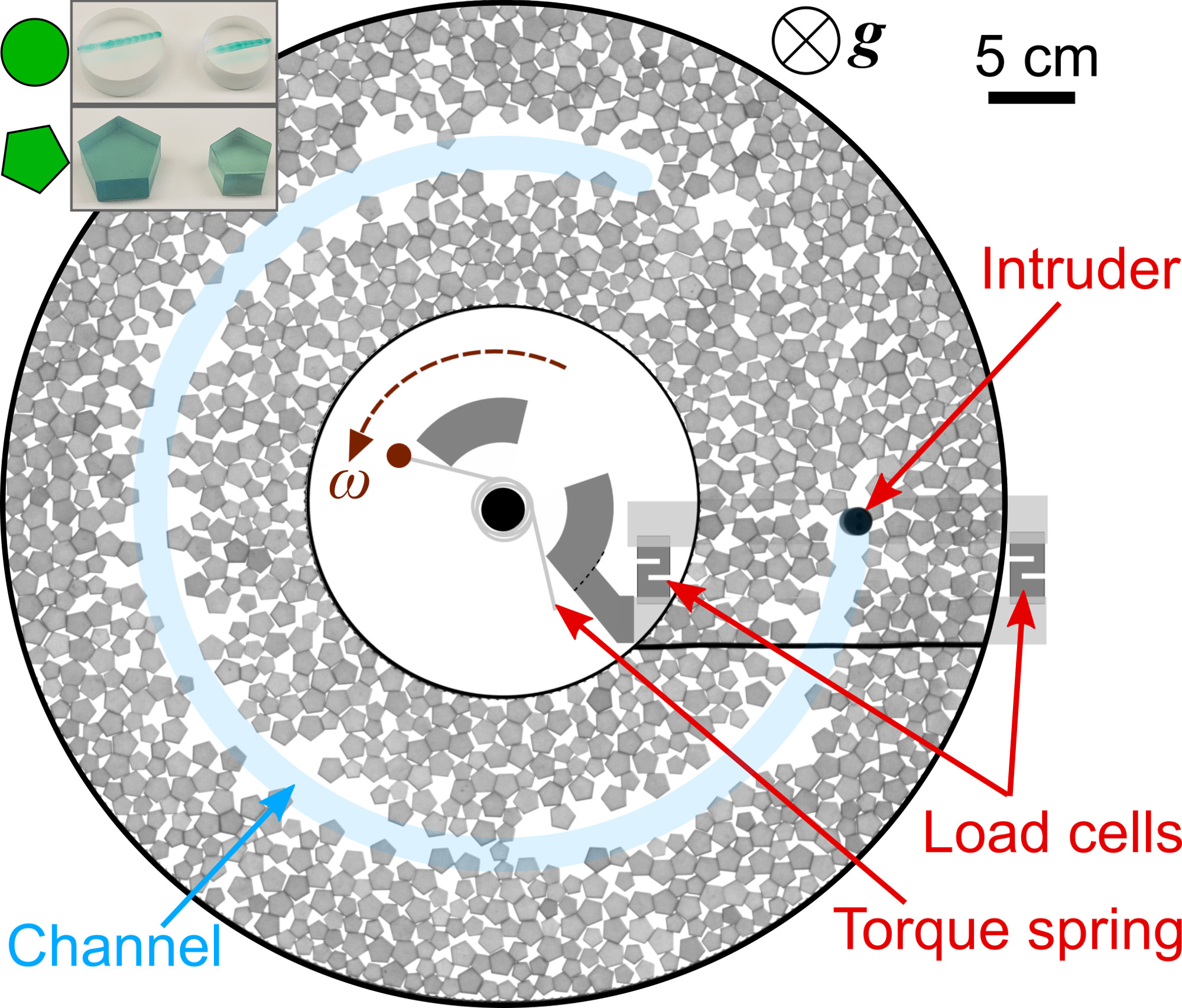}
    \caption{White light image of experimental system. The cantilever holding the intruder is driven counterclockwise at a fixed rate $\omega$ by a torque spring. Load cells in the cantilever measure the force of grains acting on the intruder in the azimuthal direction. Gravity $g$ points into the page. The channel that forms in the wake of the intruder is highlighted in blue. Inset: Photographs of disks and pentagons used in this experiment.}
    \label{fig:apparatusschematic}
\end{figure}

Approximately 1000 bidisperse polyurethane photoelastic disks or regular pentagons are set in an annular cell of fixed area, as shown in Fig.~\ref{fig:apparatusschematic}. The inner and outer walls of the cell are coated in ribbed rubber to suppress the slipping of particles at the boundaries. The number ratio of large to small particles is approximately 1:1.1 for both the disk-packing and the pentagon-packing. The diameters of the small and large disks, respectively, are $d_{\rm Ds} = 1.280 \pm 0.003$ cm and $d_{\rm Dl} = 1.602 \pm 0.003$ cm. The diameters of the circumscribing circles of the small and large regular pentagons, respectively, are $d_{\rm Ps} = 1.211 \pm 0.001$ cm and $d_{\rm Pl} = 1.605 \pm 0.001$ cm. 
The inner radius of the annulus is 11.2 cm, and its width is 17.8 cm. 
Both disks and pentagons are approximately 0.6 cm thick.  
The intergrain contact friction for all grains is $\mu = 0.7 \pm 0.1$, and the basal friction between grains and the glass base of the annulus is $\mu_{\rm BF} = 0.25 \pm 0.05$. 

The intruder, a Teflon\textsuperscript{\textregistered} rod with diameter  $\di = 1.588 \pm 0.001 \;\;\text{\rm cm} \approx d_{\rm Dl}$, is rigidly fixed to a cantilever that is attached to a post at the center of the annulus, mounted so that it suspends into the plane of the granular material and can push the particles without touching the base. The cantilever is coupled to a stepper motor by a torque spring of constant $\kappa = 0.394 \pm 0.001$ Nm/rad, which is driven by the stepper motor at a fixed rate $\omega = 0.12$ rad/s. For these parameter values, the intruder exhibits stick-slip dynamics in the granular medium when the packing fraction of grains is high enough \cite{stickslipnasuno,KozlowskiBasalFrictionPRE}. 
During sticking periods, the intruder and grains are nominally stationary, and the energy in the torque spring increases as the motor turns.  
A slip event occurs when the material fails, resulting in rapid rearrangements of grains and fast relaxation of the stress in the torque spring.  The slip event ends when the intruder encounters a jammed configuration of grains that resists further penetration. 

The cantilever holding the intruder contains load cells that record forces generated by the grains acting on the intruder. The sum of the two forces is the total azimuthal force that the grains exert on the intruder. Cameras above the system are used to both track the grains and intruder in white-light images as well as acquire photoelastic images, which allow for quantification of stresses on the grain scale \cite{photoelastictechniquedaniels,enlighteningforcechainsBares2019}. The intruder and grains are tracked via a custom correlation template matching algorithm, described in detail in Ref.~\cite{PGKoz2021} and provided in Ref.~\cite{Kozlowski2021GitHubTrackCode}. At least 99\% of grains are successfully tracked in every video frame, and grain centers are detected to within 2\% of the grain diameter.  

The number of grains is varied to change the global packing fraction $\phi$ of the medium, which is defined as the fraction of area within the annular system that is occupied by grains. We have already shown in Ref.~\cite{PGKoz2021} that pentagons resist the motion at the intruder at lower packing fractions (from $\phi = 0.61$ to  $\phi = 0.65$) than disks (from $\phi = 0.65 $ to $ \phi =  0.77$), with the strength of the pentagon-packing at $\phi=0.65$ being similar to that of the disk-packing at $\phi=0.77$. Above $\phi = 0.77$ for disks, the material quickly buckles out of the annulus; above $\phi = 0.65$ for pentagons, buckling is not common, but the material is strong enough to resist forces that exceed the limit of what the stepper motor driving the intruder can supply. 
Thus, we are not able explore higher packing fractions for disks or pentagons in this system. 
We note that $\phi = 0.77$ is close to the isotropic jamming fraction $\phi_J \approx 0.79$ measured for disks of the same material used in this experiment, but $\phi = 0.65$ is far below $\phi_J \approx 0.77$ measured for pentagons of the same material used in this experiment \cite{YiqiuSingleG2}. The uncertainties in all measured packing fractions in this work are $\delta \phi = 0.008$.

For a given $\phi$, two (or more) experimental runs are performed in which the intruder is driven thirteen revolutions around the annulus. By the end of the first four revolutions, the intruder dynamics reaches a statistically steady state. We then record data for nine revolutions, obtaining sufficient numbers of stick-slip cycles for statistical analysis within the limitations of our image acquisition system. In the steady state, as shown in Fig.~\ref{fig:apparatusschematic}, a channel is formed behind the intruder as particles, whose motions are damped by basal friction, are pushed out of its path \cite{dragforcecavityformationintruderkolb}. The intruder is able to form an open channel in the steady state for pentagons at $\phi_0 \leq 0.60$ and disks at $\phi_0 \leq 0.64$. $\phi_0$ is determined by the steady state packing fraction of grains in the regions of the annulus on either side of the intruder's path, which are lower for pentagons than for disks. In this work, we consider the rescaled packing fraction $\Delta \phi \equiv (\phi/\phi_0)-1$. 

We measure approximately 500 stick-slip cycles per $\phi$, except for low enough $\phi$ that an open channel forms or the system does not exhibit stick-slip dynamics \cite{KozlowskiBasalFrictionPRE}. In this work, $\langle \cdot \rangle$ denotes an average over the  stable structures of each sticking period. A single frame at the middle (temporally) of each sticking period is used to represent the structure of that particular period in statistical analyses.

\subsection{Grain-scale stress measurements} \label{ssec:stressmeasures}

\begin{figure}
    \centering
    \includegraphics[width=0.8\linewidth]{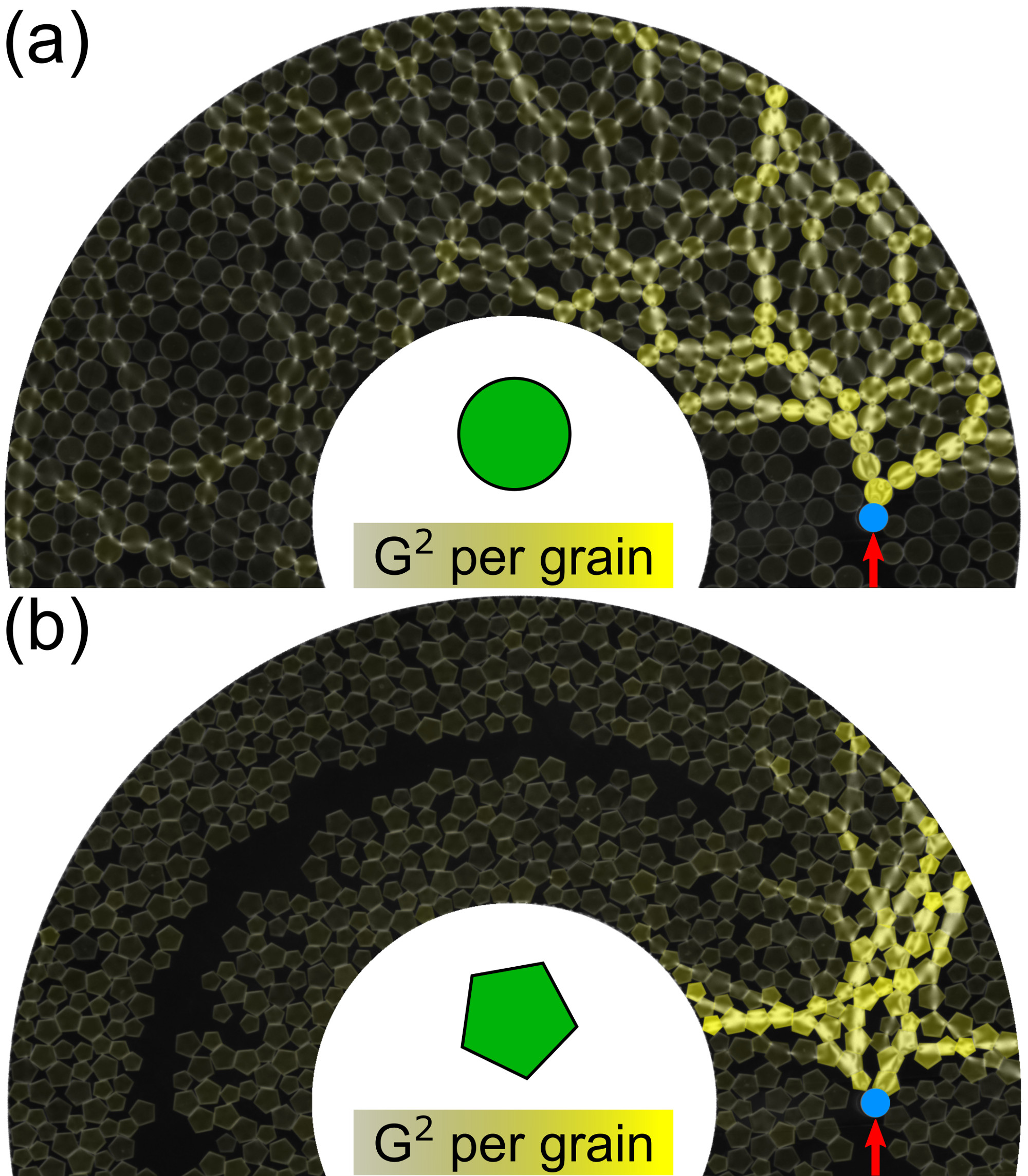}
    \caption{Dark-field photoelastic images showing the stresses (arbitrary units) of individual grains for (a) the disk-packing at $\phi = 0.77$ ($\Delta\phi = 0.20$) and (b) the pentagon-packing at $\phi=0.65$ ($\Delta\phi = 0.08$). Brighter (yellow) grains experience more stress. Both images are taken from sticking periods where the applied force from the intruder (in the counterclockwise direction, denoted by the red arrow) is $\sim 2.4 \text{ N}$; the yellow color scale is the same for both images.}
    \label{fig:g2pergrainimage}
\end{figure}

A dark-field polariscope is used to visualize grain-scale stresses  \cite{photoelastictechniquedaniels,enlighteningforcechainsBares2019}. 
The gradient-squared $G^2$ of the intensity in a photoelastic image scales with the magnitude of forces applied at point contacts \cite{Geng2003greensfunction}. Although the nature of the fringe patterns is complicated by the presence of edge-edge contacts in angular grain packings \cite{yiqiushape}, 
we have confirmed the efficacy of $G^2_i$ as a semi-quantitative estimate of stress with our pentagonal grains by applying known forces to the grains using various combinations of vertex-edge and edge-edge contacts, according to the methods established in Ref.~\cite{yiqiushape}.  
Figure~\ref{fig:g2pergrainimage} shows photoelastic images overlaid with semi-transparent grain images artificially colored according to extracted values of  $G^2_i$, brighter yellow indicating larger $G^2_i$.   The match between the photoelastic pattern and the $G^2_i$ pattern demonstrates the qualitative effectiveness of this measure even for pentagons experiencing edge-edge contacts.

\begin{figure}
    \centering
    \includegraphics[width=0.6\linewidth]{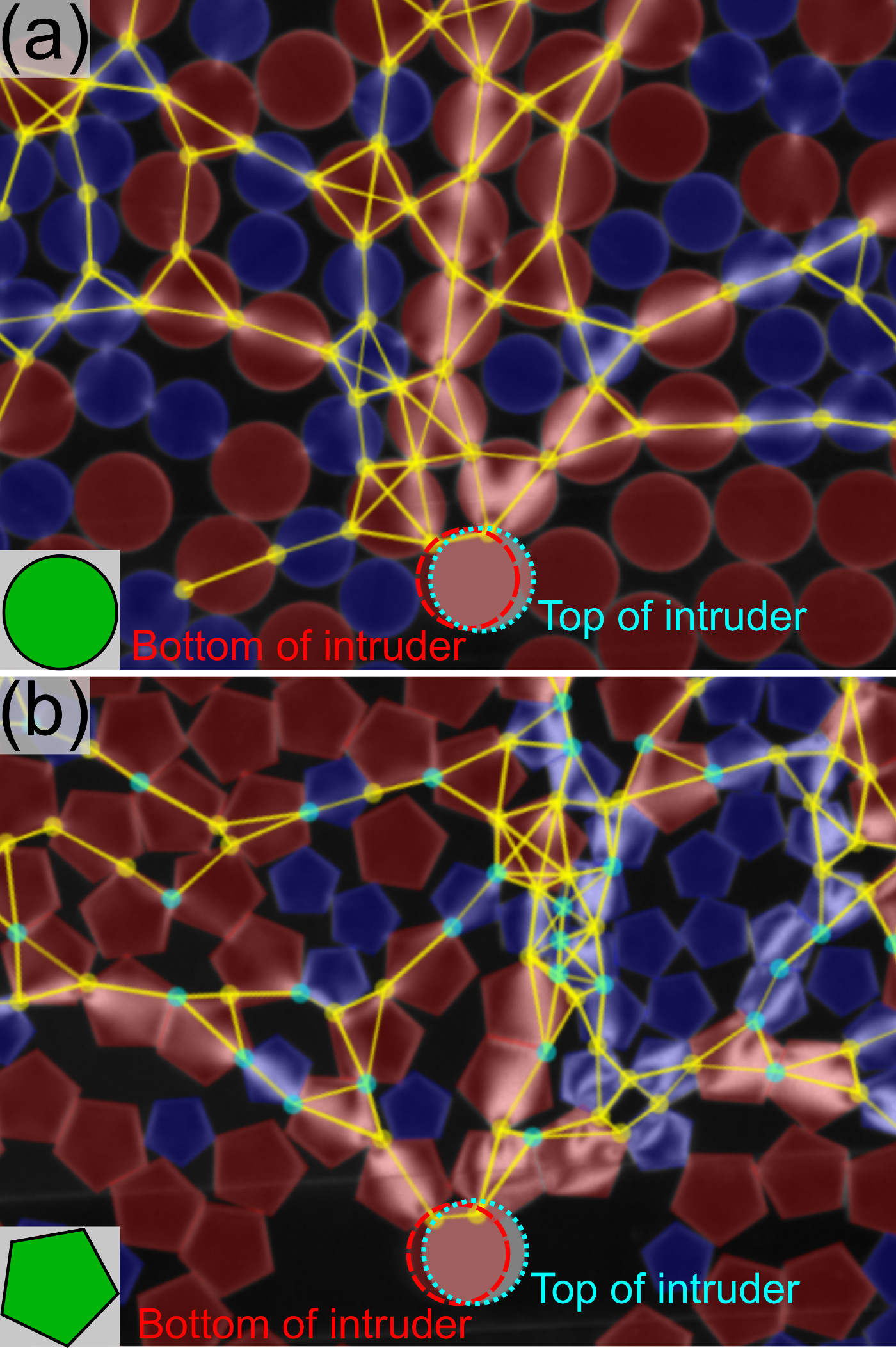}
    \caption{Photoelastic images of (a) a sample disk-packing and (b) a sample pentagon-packing overlaid with semi-transparent detected grains (red = large grain, blue = small grain) and detected force-bearing contact points (yellow = disk contacts or vertex-edge contacts for pentagons, cyan = edge-edge contacts). The top of the intruder (above the imaging plane of the granular material, see Sec.~\ref{sec:expmethods} and Ref.~\cite{KozlowskiBasalFrictionPRE,PGKoz2021}) is indicated by a cyan dotted circle, and the bottom, which contacts the granular material, by a red dashed circle.}
    \label{fig:contactdetection}
\end{figure}

We also analyze the network of \textit{force-bearing} contacts; sample photoelastic images with tracked grains and detected contacts are shown in Fig.~\ref{fig:contactdetection}. First, \textit{geometric contacts} are detected between grains that are closer than a threshold separation distance. The photoelastic image is then checked in a small region around the geometric contact point for intensities exceeding an intensity threshold to determine whether the contact is force-bearing. Edge-edge contacts are defined as contacts for which the relative angle between the two pentagons is smaller than a threshold $8^{\circ}$. These contacts are shown as cyan dots in Fig.~\ref{fig:contactdetection}(b), with the contact point defined as the location of highest local $G^2$ along the contacting edges. With force-bearing contacts detected, we can quantify structural properties of force chains, the mesoscale structures that bear the greatest stress from the intruder \cite{Cates1998JammingChainsFragile,Radjai1998BimodalStressTransmission,howellG2stressfluctuations,TordPREforcechains}.

\begin{figure*}[h]
    \centering
    \includegraphics[width=0.9\linewidth]{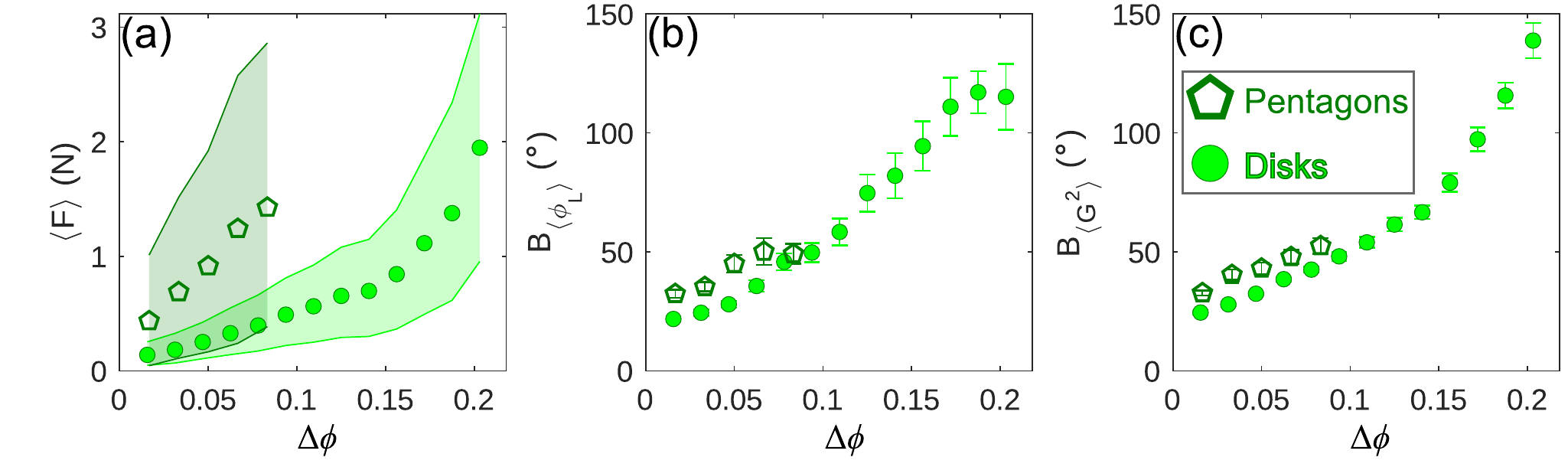}
    \caption{(a) The average force $\langle F\rangle$ exerted on the intruder by grains  as the intruder is driven through the pentagon-packing (orange pentagons) and disk-packing (purple circles) as a function of $\Delta \phi = \phi/\phi_0-1$.  The range of $\langle F\rangle$ is from the lower 10\% cutoff to the upper 90\% cutoff of the intruder's force distribution.  (b) Decay constant $B_{\langle\phi_L\rangle}$ from Eqn.~\ref{eqn:phildecay} of local packing fraction along the annulus from the intruder. (c) Decay constant $B_{\langle G^2 \rangle}$ from Eqn.~\ref{eqn:G2decay} of stress along the annulus from the intruder. Error bars for (b) and (c) are 95\% confidence intervals of the decay constants from the exponential fits Eqn.~\ref{eqn:phildecay} and Eqn.~\ref{eqn:G2decay}.}
    \label{fig:rescalephiplots}
\end{figure*}

\section{Stresses in the granular material}
\label{sec:stressprop}

With grain stresses obtained from photoelastic images and grain positions and orientations from white light images, we analyze the spatial distribution of grains and stresses during sticking periods. We first quantify the spatial extent of the cluster of grains in front of the intruder (or the spatial extent of the channel behind the intruder, see Fig.~\ref{fig:apparatusschematic}) and then quantify similarly the spatial extent of stresses in the granular material along the annulus.

\subsection{Spatial packing fraction and stress distributions}
\label{ssec:annularstress}

\begin{figure}
    \centering
    \includegraphics[width=1\linewidth]{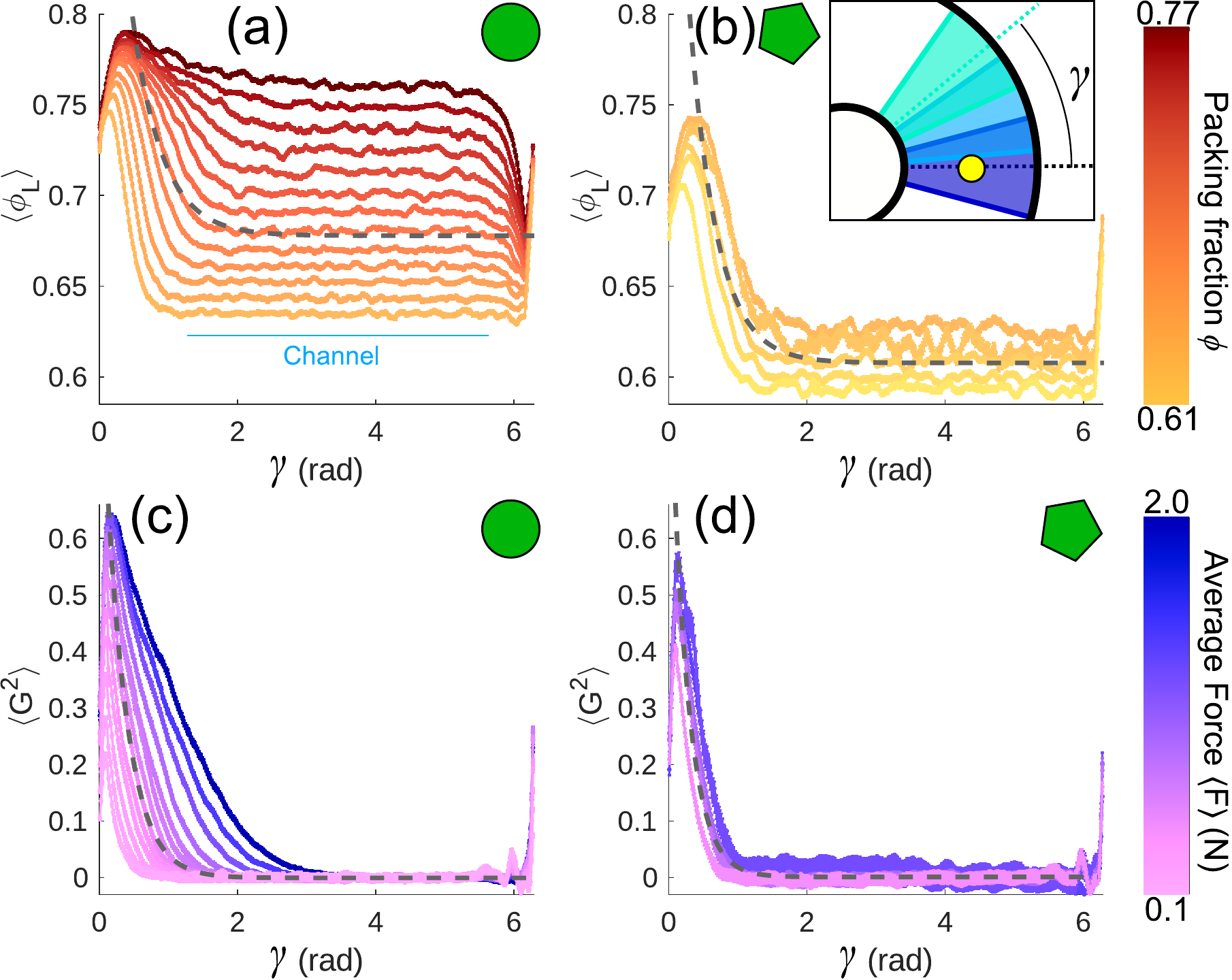}
    \caption{ (a,b) Average local packing fraction $\langle \phi_L\rangle$ measured in wedges, starting centered at the intruder $\gamma=0$ and sampled around the entire annulus, (a) for the disk-packing and (b) for the pentagon-packing. (b) schematically demonstrates three wedges at consecutive angles $\gamma$ used to sample local packing fraction. The color scale corresponds to the global packing fraction $\phi$. $\langle \phi_L\rangle$ is constant in regions with the channel formed behind the intruder. (c,d) Average normalized gradient-squared per grain measured in the same sampling wedges as in (a,b). The color scale corresponds to the average force the granular material exerts on the intruder for each global packing fraction (see Fig.~\ref{fig:rescalephiplots}(a)). Sample exponential decay fits are shown (gray dashed curves) for disks at $\phi=0.70$ and pentagons at $\phi = 0.63$.}
    \label{fig:localphiG2angularwedge}
\end{figure}

Figure~\ref{fig:rescalephiplots}(a) shows the average force and width of the distribution of forces exerted on the intruder as a function of the rescaled packing fraction $\Delta \phi \equiv \phi/\phi_0-1$, where $\phi_0=0.60$ for the pentagon-packing and $0.64$ for the disk-packing (Sec.~\ref{ssec:system}). The pentagon-packing produces comparable average forces on the intruder at significantly lower $\Delta \phi$ (and $\phi$). In the steady state, the cluster of grains in front of the intruder is also significantly smaller. We measure the extent of this high-density cluster of grains, shown in Fig.~\ref{fig:localphiG2angularwedge}(a,b), by computing the average local packing fraction $\langle \phi_L\rangle (\gamma)$ in a wedge along the annulus at angle $\gamma$ with respect to the intruder (inset of Fig.~\ref{fig:localphiG2angularwedge}(b)). The sampling wedge is $8^{\circ}$ wide ($\sim d_{\rm Dl}$ wide at the inner boundary) and centered on $\gamma$, spanning the full radial range of the annulus. The wedge is rotated in increments of $1^{\circ}$ to smoothly sample the local grain density around the annulus. The angle of the intruder is defined as $\gamma=0^{\circ}$.

Figure~\ref{fig:localphiG2angularwedge}(a,b) demonstrates that $\langle \phi_L\rangle(\gamma)$ is largest in front of the intruder and decays to a $\phi$-dependent constant that corresponds to the channel left in the wake of the intruder. (Note that since the sampling wedge is centered on $\gamma$, $\langle \phi_L \rangle (\gamma=0)$ centered on the intruder is low because the wedge includes part of the channel directly behind the intruder.) For both disks (a) and pentagons (b) $\langle \phi_L \rangle$ decays to a constant more rapidly as $\phi$ decreases. We quantify this decay through an exponential fit 
\begin{equation} \label{eqn:phildecay}
    \langle \phi_L\rangle (\gamma) = A\exp\bigg(\frac{-2\gamma}{B_{\langle\phi_L\rangle}}\bigg)+D
\end{equation}
starting at the peak in $\langle \phi_L \rangle$. Fig.~\ref{fig:rescalephiplots}(b) shows decay constant $B_{\langle\phi_L\rangle}$, an estimate of the spatial extent of packing fraction, for both disks and pentagons as a function of $\Delta \phi$. Given that the highest accessible $\phi$ for pentagons is low compared with disks, the extent of the high-density cluster is much more limited. The decay of $\langle \phi_L \rangle$ is comparable for disks and pentagons at low $\Delta \phi$.

We next measure the spatial extent of stress in the packing with respect to the angle  of the intruder, averaged over all sticking periods. First, for a given sticking period, the average $G^2$ for each grain  $G^2_i$ is computed. Next, the average of $G^2_i$ for all grains whose centers fall in the sampling wedge (same wedge as described above, shown in Fig.~\ref{fig:localphiG2angularwedge}(b)) centered on $\gamma$, $G^2(\gamma)$, is obtained. Lastly, $G^2(\gamma)$ is normalized by the maximum value of $G^2$ along the channel. $\langle G^2\rangle (\gamma)$, a semi-quantitative estimate of the stress (Sec.~\ref{ssec:stressmeasures}) along the annulus shown in Fig.~\ref{fig:localphiG2angularwedge}(c,d), is then obtained by averaging the normalized angular distribution over all sticking periods. 

As with $\langle \phi_L \rangle (\gamma)$, $\langle G^2\rangle (\gamma)$ peaks near the intruder and decays along the annulus at a rate that is faster for lower $\phi$. The spatial extent of the stress field was quantified through an exponential fit of the form 
\begin{equation} \label{eqn:G2decay}
    \langle G^2 \rangle (\gamma) = A'\exp\bigg(\frac{-2\gamma}{B_{\langle G^2 \rangle}}\bigg)+D'
\end{equation}
starting at the peak in $\langle G^2 \rangle$. Fig.~\ref{fig:rescalephiplots}(c) shows the decay constant $B_{\langle G^2 \rangle}$.

The decay constants associated with local packing fraction and local stress are similar to each other, which implies that most stresses are transmitted in the dense cluster of grains in front of the intruder. For the densest packing of pentagons, the extent of stresses along the annulus roughly corresponds to the extent of the dense cluster. As with the spatial extent of $\langle\phi_L\rangle$, the extent of stresses collapses for pentagons and disks when $\phi$ is rescaled by $\phi_0$. As indicated by the color scale used in Fig.~\ref{fig:localphiG2angularwedge}(c,d), however, the pentagons exert substantially larger forces on the intruder even when stresses are confined to a smaller region. 

\subsection{Back-bending of force chains}
\label{ssec:backbending}

\begin{figure}
    \centering
    \includegraphics[width=1\linewidth]{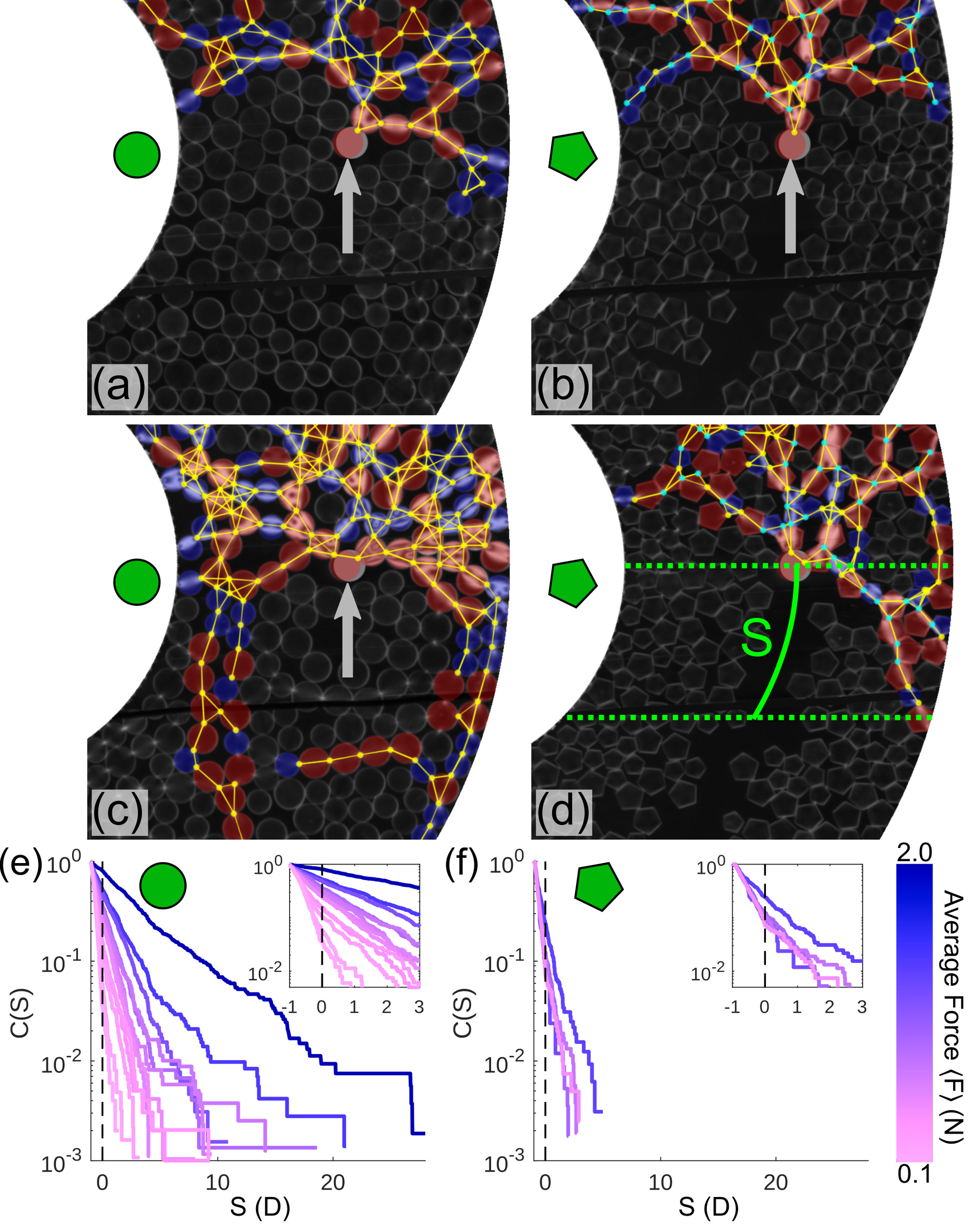}
    \caption{(a-d) Sample images of back-bending force chains for (a,c) disks at $\phi = 0.77$ and (b,d) pentagons at $\phi = 0.65$. The gray arrow indicates the direction in which the intruder is applying force. The arclength $S$ describing the amount of back-bending for a given sticking period is defined schematically in (d). (d) is one of the few rare structures for pentagons with a large $S$. (e,f) Complementary cumulative distribution function $C(S)$ for disks (e) and pentagons (f) with varying $\phi$. The vertical dotted line corresponds to $S=0$; $C(S<0)$ indicates sticking periods in which force chains only propagate forward from the intruder, as in the sample image in (b).}
    \label{fig:backbendingstresschains}
\end{figure}

In Sec.~\ref{ssec:annularstress}, it was noted that force chains in disk-packings wrap around the annular system more than in pentagon-packings with comparable applied forces from the intruder. We also observe that force chains  propagating behind the intruder (that is, opposite of the direction of applied force from the intruder) in stable sticking periods occur more frequently for disks than for pentagons. Examples of this behavior are shown in photoelastic images of four sample sticking periods for disks and pentagons in Fig.~\ref{fig:backbendingstresschains}(a-d), with detected contact points and lines of grains topologically connected with the intruder drawn over the images. As shown in Fig.~\ref{fig:backbendingstresschains}(d), the back-bending extent $S$ is defined as the arclength from the intruder (where angle is measured with respect to the center of the system) of the furthest grain behind the intruder that is topologically connected with the intruder in the force network.  

\begin{figure}
    \centering
    \includegraphics[width=1\linewidth]{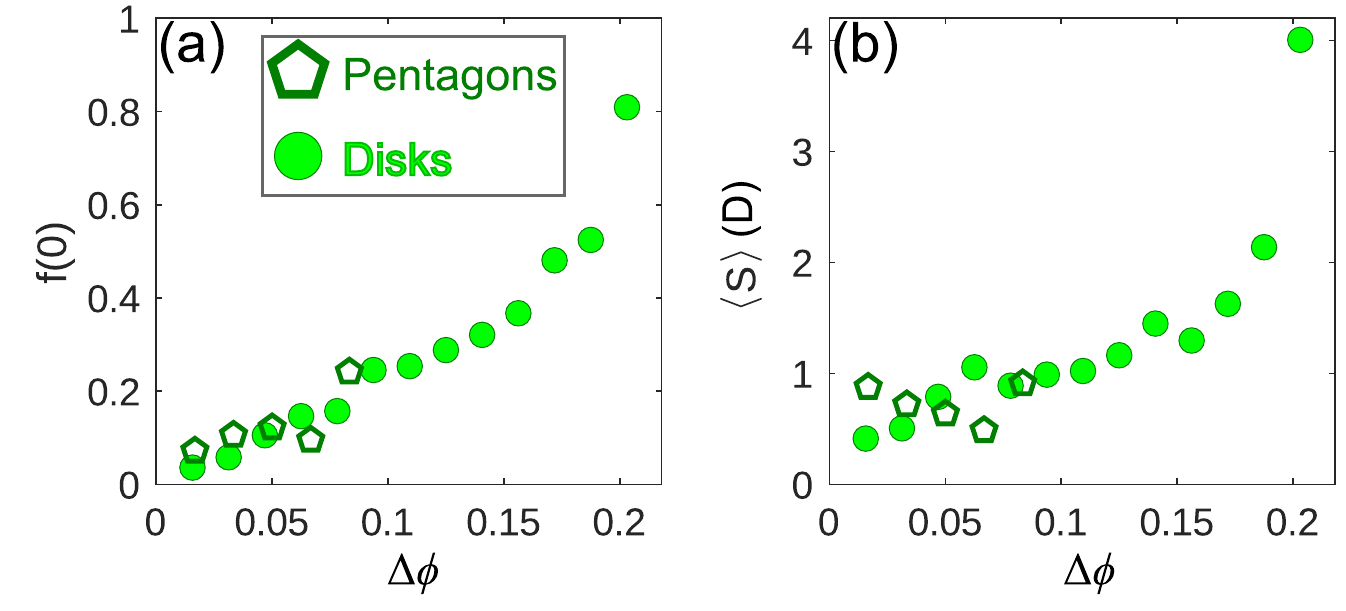}
    \caption{(a) Fraction $f$ of sticking periods that feature back-bending force chains ($S>0$) for pentagons (orange) and disks (purple) with $\Delta \phi = \phi/\phi_0-1$. (b) Average of back-bending extent $\langle S \rangle$, excluding events that do not feature back-bending force chains.}
    \label{fig:backbendstatsphi}
\end{figure}

Figure~\ref{fig:backbendingstresschains}(e,f) shows complementary cumulative distribution functions $C(S)$ for pentagons and disks for different $\phi$'s. Note that a substantial fraction of stable structures do not feature back-bending chains at low $\phi$ for disks and at all accessible $\phi$ for pentagons; these structures correspond to $-\di<S<0$. The distribution tail broadens for increasing $\phi$, and at the largest packing fractions attainable for disk- and pentagon-packings, disks demonstrate a much larger probability to propagate forces behind the intruder, and do so over much greater distances. At the highest $\phi$ for pentagons, a small number of larger $S$ sticking periods occur; Fig.~\ref{fig:backbendingstresschains}(d) shows one of these rare occurrences. Figure~\ref{fig:backbendstatsphi} summarizes these differences by showing the fraction of sticking periods that feature back-bending force chains and their  average extent $\langle S\rangle$. Though the resistance of the pentagon-packing is greater, the statistics related to back-bending are similar for disks and pentagons at comparable $\Delta \phi$.

\subsection{Edge-edge contacts}

\begin{figure}
    \centering
    \includegraphics[width=0.8\linewidth]{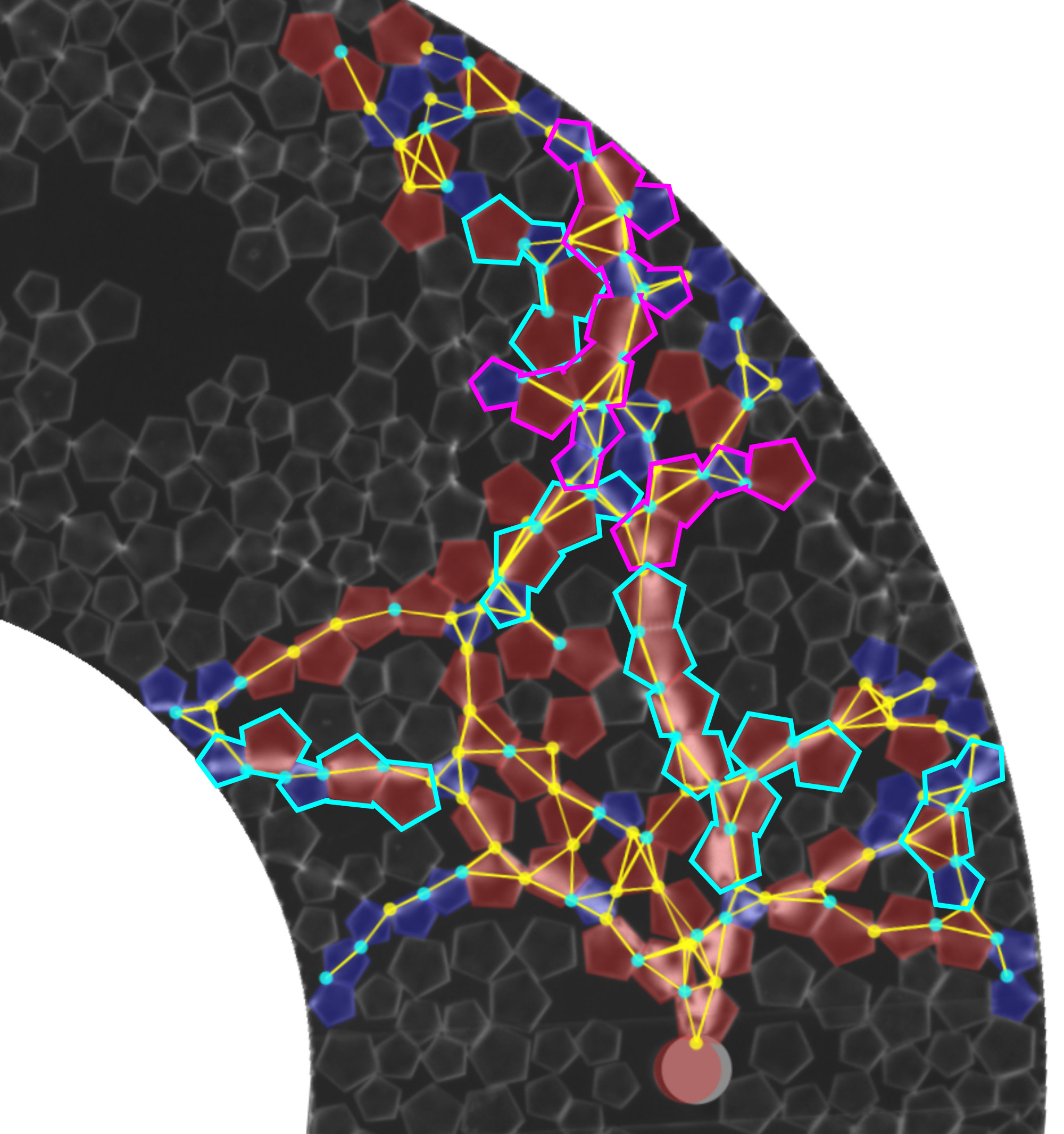}
    \caption{Zoom-in of sample photoelastic image from Fig.~\ref{fig:backbendingstresschains}(b) during a sticking period for pentagons at $\phi = 0.65$. Sequences of detected edge-edge contacts featuring 4 or more grains are drawn around the edges in cyan (or purple to distinguish adjacent sequences).}
    \label{fig:sampleedgeedge}
\end{figure}

Photoelastic images from experiments give the  impression that pentagons propagate stronger forces along columnar sequences of edge-edge contacts, as illustrated in   Figure~\ref{fig:sampleedgeedge}, where edge-edge contact chains of four or more grains are highlighted along their perimeters.  (Similar configurations can be seen in Figs.~\ref{fig:g2pergrainimage}, \ref{fig:contactdetection}, and \ref{fig:backbendingstresschains}.)   These sequences of edge-edge contacts are indeed common in the pentagon packings, and the rotational constraints associated with such contacts may allow them to support the strongest stresses in the packing \cite{pentagonalforcetransmissionazema,yiqiushape}. In our system, however, quantitative determination of forces at edge-edge contacts remains elusive, as mentioned in Sec.~\ref{sec:expmethods}.  It is thus difficult to collect reliable data on the correlations between contact geometries and force magnitudes. Numerical simulation studies, in which all contact forces are precisely known, as well as extensions of photoelastic contact force extraction to polygonal particles, could lend important insights into the role of edge-edge contacts in determining the structure of strong force chains and hence into the differences in the stick-slip behaviors observed in pentagon and disk packings.

\section{Conclusions}\label{sec:conclusion}

We have demonstrated  that, compared to disk-packings, pentagon-packings impede the progress of a single-grain intruder with a smaller cluster of grains in front, a smaller spatial extent of stresses, and fewer back-bending forces. Nonetheless, these pentagon-packings are able to  resist the  intruder at significantly lower global packing fractions than are required in disk-packings. The scaling collapse of the cluster extent, stress extent, and back-bending force chain properties between disk- and pentagon-packings implies that the two granular materials transmit stresses similarly from the local load to the boundaries when the packing density profiles are similar.  However, the magnitude of stresses that can be sustained at similar profiles is significantly different; for a given density profile, pentagons can exert much larger forces before yielding than disks can. Given that the grains in all of our experiments have similar sizes and identical elastic moduli, we conclude that grain angularity has a strong influence on the resistance of a granular material to a local intruding load. These findings, along with the flow patterns presented in Ref.~\cite{PGKoz2021}, all form a consistent description of the behavior of locally loaded pentagons in this system. 

During slip events, pentagons primarily get pushed forward, filling in the open channel created by a previous pass of the intruder. Disks, on the other hand, circulate around the intruder and can fill in the channel behind it\cite{PGKoz2021}. This circulation of disks during slip events could explain the prevalence of back-bending force chains for the disk-packing -- these structures are dynamically formed during slip events, finding support from the annulus walls as well as (weak) basal friction forces. Since pentagons tend to be pushed forward during slips, such structures are rarely formed. We note here as well that the presence of back-bending force chains has been predicted in a recently developed theory of elasticity of isostatic granular media  \cite{RafiBlumenfeldBackbending}, even with quasi-static, non-inertial loading. Our findings suggest that the rotational constraints introduced by edge-edge contacts between pentagons modify the predictions of this theory, possibly creating a larger critical radius of curvature from the loading source. The current system's limited channel width of $\sim 15$ small grains, however, prevents a thorough investigation of the curvature of the force chains. 
Whether backwards curvature of pentagon-packing force chains will cause more flow around the intruder for sufficiently large channel widths remain an open question.  If so, it is possible that the strength of the pentagon-packing might resemble more closely the strength of the disk-packing at comparable $\phi$.

Our results provide evidence that granular media with non-circular features, in this case flat edges and vertices, transmit stresses differently from a localized load than idealized (circular) grains. This finding provides more evidence that numerical simulations and controlled experiments will better capture meso- and macro-scale behaviors of real granular materials in nature when they account for micro-scale interactions arising from changes to grain shape.

\section*{Author contributions}
R. Kozlowski contributed to data curation, formal analysis, investigation, methodology, validation, visualization, and writing of the original draft and edited versions.
H. Zheng contributed to methodology, conceptualization, supervision, and writing edited versions.
K. E. Daniels contributed to formal analysis, investigation, methodology, supervision, and writing of the original draft and edited versions.
J. E. S. Socolar contributed to formal analysis, funding acquisition, methodology, investigation, supervision, and writing of the original draft and edited versions.

\section*{Conflicts of interest}
There are no conflicts to declare.

\section*{Acknowledgements}
This work was supported by the US Army Research Office through grant W911NF1810184. 


\balance



\providecommand*{\mcitethebibliography}{\thebibliography}
\csname @ifundefined\endcsname{endmcitethebibliography}
{\let\endmcitethebibliography\endthebibliography}{}

\bibliographystyle{rsc.bst}

\end{document}